\documentclass[aps,twocolumn,prl,superscriptaddress]{revtex4-1}
\usepackage{amsmath}
\usepackage{amssymb}
\usepackage{graphicx}
\usepackage{color}

\newcommand{\comment}[2]{\begingroup\color{red}\em(#2 ---#1)\endgroup}
\DeclareMathOperator{\Tr}{Tr}

\DeclareMathOperator{\saw}{saw}
\renewcommand{\Re}{\mathop{\mathrm{Re}}}
\renewcommand{\Im}{\mathop{\mathrm{Im}}}
\newcommand{\dd}[1]{\mathop{\mathrm{d}#1}}

\newcommand{\avg}[1]{\langle{#1}\rangle}

\newcommand{\tz}{\hat{\tau}_{3}}

\renewcommand{\comment}[2]{}

\begin{document}
\title{Linear AC Response of Diffusive SNS Junctions}

\author{Pauli Virtanen}
\affiliation{Institute for Theoretical Physics and Astrophysics,
  University of W\"urzburg, D-97074 W\"urzburg, Germany}
\affiliation{Low Temperature Laboratory, Aalto University,
  P.O. Box 15100, FI-00076 AALTO, Finland}

\author{F. Sebasti\'an Bergeret}
\affiliation{Centro de F\'{\i}sica de Materiales (CFM), Centro Mixto
  CSIC-UPV/EHU, Edificio Korta, Avenida de Tolosa 72, E-20018 San Sebasti\'an,
  Spain}
\affiliation{Donostia International Physics Center (DIPC),
  Manuel de Lardizbal 4, E-20018 San Sebasti\'an, Spain}

\author{Juan Carlos Cuevas}
\affiliation{Departamento de F\'{\i}sica Te\'orica de la Materia
  Condensada, Universidad Aut\'onoma de Madrid, E-28049 Madrid, Spain}

\author{Tero T. Heikkil\"a}
\affiliation{Low Temperature Laboratory, Aalto University,
  P.O. Box 15100, FI-00076 AALTO, Finland}

\date{\today}

\begin{abstract}
  We explore the behavior of the ac admittance of superconductor-normal
  metal-superconductor (SNS) junctions as the phase difference
  of the order parameters between the superconductors is varied. We find
  three characteristic regimes, defined by comparing the driving
  frequency $\omega$ to the inelastic scattering rate $\Gamma$ and the
  Thouless energy $E_T$ of the junction (typically $\hbar\Gamma \ll
  E_T$). Only in the first regime $\omega \ll \Gamma$ the usual
  picture of the kinetic inductance holds. We show that the ac
  admittance can be used to directly access some of the
  characteristic quantities of the SNS junctions, in particular the
  phase dependent energy minigap and the typically phase dependent
  inelastic scattering rate. Our results partially explain the
  recent measurements of the linear response properties of SNS
  Superconducting Quantum Interference Devices (SQUIDs) and predict
  a number of new effects.
\end{abstract}

\maketitle

The frequency dependent susceptibility typically reveals information
about the internal dynamics of the studied systems. In the electronic
case, this susceptibility is more often measured as admittance, whose
frequency dependence in semiclassical models for bulk metals is due to
scattering and appears for frequencies exceeding some tens of THz
\cite{ashcroft1976-ssp}, and in wires it is dictated by stray
capacitances and geometric inductance.  Understanding the
frequency-dependent response is moreover of importance for
high-frequency devices.

In superconductors or superconducting tunnel junctions (SIS), the
admittance at low frequencies is dominated by the superconducting kinetic
inductance \cite{tinkham1996-its}. For SIS junctions, this Josephson
inductance is related to the supercurrent $I_S(\varphi)$ through the
junction, which depends on the superconducting phase difference
$\varphi$:
\begin{equation}
  L_J(\varphi)^{-1}=\frac{2e}{\hbar} \partial_\varphi I_S(\varphi)
  \,.
  \label{eq:josephsoninductance}
\end{equation}
This relation allows characterizing the current-phase relation of
Josephson junctions via measurements of their ac admittance
\cite{golubov2004-cri}. The remaining dissipative part of the
admittance is due to quasiparticles, is proportional to
$\exp[-|\Delta-\hbar \omega|/(k_B T)]$ \cite{tucker1985-qda}, and
is important only for the high frequencies of the order of the
superconducting gap $\Delta$ or temperatures $T$ close to the critical
temperature.

For other types of Josephson junctions than SIS the admittance,
however, can deviate from the above simple picture. In this Letter we
show how combining normal metal wires (N) and superconductors (S) into
SNS weak links results into an admittance that entails characteristics
of the inelastic scattering rates and the inverse diffusion times
through the structure. At frequencies of the order or larger than the
inelastic scattering rate $\Gamma$, the simple kinetic inductance
picture has to be revised to include non-adiabatic effects associated
with the dynamics of the electron distribution.  The dissipative response,
describing microwave absorption, is moreover finite for
temperatures or frequencies exceeding the phase-dependent minigap
in the spectrum of excitations inside the junction. It probes the
density of states in the junction, and is related to the physics of
stimulation and suppression of the supercurrent
\cite{warlaumont1979-mpe,*fuechsle2009-eom,*aslamazov1982-ssb,virtanen2010-tom}.

Here, we study diffusive SNS junctions, whose length $L$ is longer
than the superconducting coherence length $\xi_0=\sqrt{\hbar
  D/\Delta}$, where $D$ is the diffusion constant. The proximity
effect induces a gap in the density of states inside the
normal metal, $E_g(\varphi=0)=3.12 E_T\ll{}\Delta$, where $E_T=\hbar
D/L^2$ is the inverse diffusion time. The minigap depends on the phase
difference approximately as $\cos(\varphi/2)$ and vanishes for
$\varphi=\pi$.  The coupling to the electric field is modeled by
assuming an ac bias voltage, $V(t)=\delta V \sin(\omega t)$, which
induces an oscillating phase difference $\varphi(t)=\varphi+\delta\phi
\cos(\omega t)$ across the junction.

To find quantitative results, we describe the SNS junction dynamics
with the Keldysh-Usadel
equation~\cite{usadel1970-gde,*belzig1999-qgf}, used also in
Ref.~\cite{virtanen2010-tom}. In this approach, physical quantities
are obtained from the Retarded, Advanced and Keldysh Green's functions
$\hat{g}^{R/A/K}(E,E')$, which depend on two energy arguments.  These
functions are matrices in the Nambu (electron-hole) space, and the
Keldysh (K) part can be parameterized in terms of an electron
distribution function matrix $\hat{h}(E,E')=h_L(E,E')+h_T(E,E')\tz$:
$\hat{g}^K=\hat{g}^R\hat{h}-\hat{h}\hat{g}^A$, where matrix products
involve also convolutions,
$(\hat{B}\hat{C})(E,E')=\int_{-\infty}^\infty\frac{\dd{E_1}}{2\pi}\hat{B}(E,E_1)\hat{C}(E_1,E')$.
In the presence of the harmonic drive, these functions can be written
in a matrix representation
$\hat{g}_{n,m}(E)\hat{=}\hat{g}(E+n\hbar\omega,E+m\hbar\omega)$
\cite{cuevas2006-pea} that reduces convolutions to matrix products.

The ac admittance is
$Y(\omega)=\frac{2}{i\omega\delta\phi} I(\omega)$, for
a linear-response drive $\delta\phi\ll2\pi$. The ac current harmonic is
$I(\omega)=\frac{\sigma_N S}{4}\int_{-\infty}^\infty\dd{E}{\rm Tr} \hat{\tau}_3 \hat{j}^K_{01}(E)$,
where $\sigma_N$ and
$S$ are the normal-state conductivity and the cross section of the
junction, and the
Keldysh current is $\hat{j}^K=\hat{g}^R\hat{\partial}_x
\hat{g}^K+\hat{g}^K \hat{\partial}_x \hat{g}^A$.
Here $x$ is the position in the junction, and
$\hat{\partial}_x\hat{B}=\partial_x\hat{B}-i[A\hat{\tau}_3, \hat{B}]$
the gauge-covariant gradient containing the vector potential
$A(E,E')=A(E)\delta(E-E')$.

In the above approach, the admittance splits naturally into three
gauge-invariant parts, $Y=Y_{\rm sc}+Y_{\rm dy}+Y_{\rm qp}$, where
(hereafter, $\hbar=e=k_B=1$)
\begin{subequations}
\begin{align}
  Y_{\rm sc}&=
  \frac{\sigma_N S}{2i\omega\delta\phi}
  \int_{-\infty}^\infty\dd{E}\Tr\{\tz(\hat j_{01}^R h_{11}-h_{00}\hat
  j_{01}^A)\}
  \,,
  \label{eq:sccurrent}
  \\
  Y_{\rm dy}&=\frac{\sigma_N S}{2i\omega\delta\phi}
  \int_{-\infty}^\infty\dd{E}\Tr\{\tz(\hat j^R_{00}
  \hat{h}_{01}-\hat{h}_{01}\hat j^A_{11})\}
  \,,
  \label{eq:dycurrent}
  \\
  Y_{\rm qp}&=
  \frac{\sigma_N S}{2i\omega\delta\phi}
  \int_{-\infty}^\infty\dd{E}\Tr\Bigl\{
  (1-\tz\hat g_{11}^A\tz\hat g_{00}^R)
  \label{eq:qpcurrent}
  \\\notag&\quad
  \times
  [
  \tz\partial_x\hat{h}_{01}
  -
  \frac{iA_0}{2}
  (h_{11}-h_{00})
  ]
  \Bigr\}
  \,.
\end{align}
\label{eq:admittancecontributions}
\end{subequations}
Here $\hat j^{R/A}=\hat{g}^{R/A}\hat{\partial}_x\hat{g}^{R/A}$
describe spectral (super)currents,
$h_{00}(E)=h_{11}(E-\omega)=\tanh(E/2T)$ the equilibrium electron
distribution, and $\hat{h}_{01}$ its time-dependent part. The
contributions describe (a) the ac supercurrent, (b) effect of the
dynamic variation of the populations of the Andreev levels, and (c) the
quasiparticle current driven directly by the field.  Below, we mostly
work in a gauge in which the electric field is contained in the vector
potential, $A(t)=A_0 \cos(\omega t)$,
$A_0=-\delta\phi/(2L)$. Requiring charge neutrality leads to a
finite position dependent scalar potential, but our numerics indicate
that this can be disregarded for $\omega\lesssim{}10E_T$.

\begin{figure}
  \includegraphics{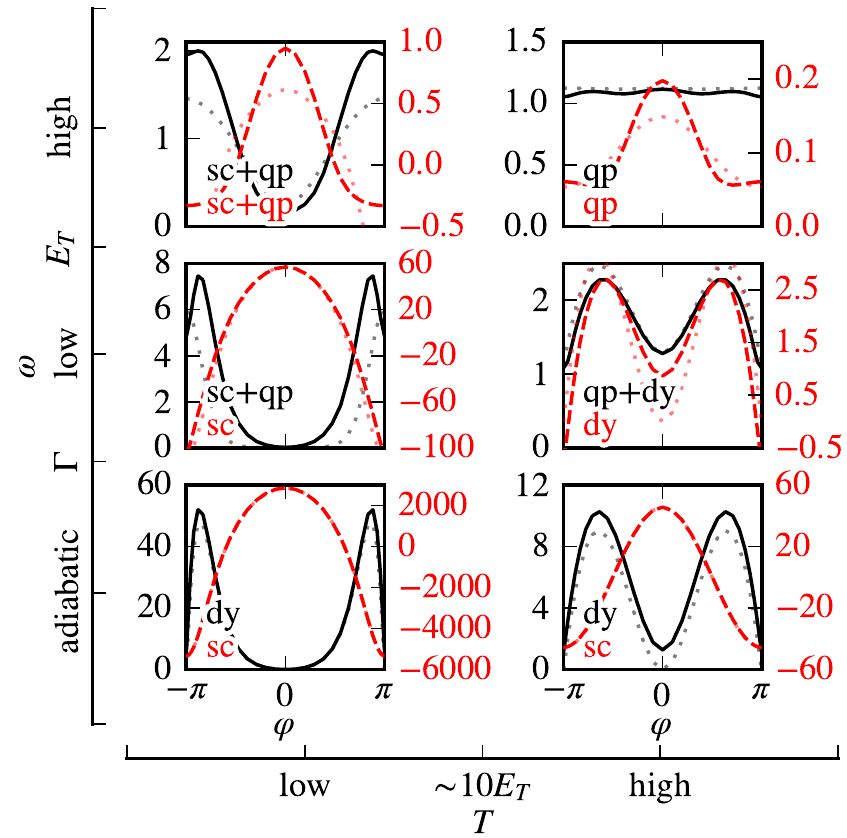}
  \caption{
    \label{fig:phasedepregimes}
    (Color online):
    Phase dependence of the admittance $Y$ in units of the normal state conductance $G_N$
    (dissipative part: solid, reactive part: dashed),
    in different regimes of interest.
    The ``low'' temperature results have been calculated for
    $T=E_T$, the ``high'' temperature results for $T=16E_T$.
    The adiabatic frequency is $\omega=E_T/200 \ll \Gamma=E_T/20$,
    the low $\omega=E_T/4\gg\Gamma$, and the high frequency is
    $\omega = 10 E_T$.
    Dotted lines show the contribution from dominant parts,
    $Y_{\rm sc}$ [Eqs.~\eqref{eq:josephsoninductance}, \eqref{eq:sc-highfreq}],
    $Y_{\rm dy}$ [Eq.~\eqref{eq:dynamic}],
    $Y_{\rm qp}$ [Eqs.~\eqref{eq:dissipative}, \eqref{eq:iqp-highfreq}],
    indicated in the lower left corners.
    The adiabatic frequency results are obtained from analytical approximations,
    the others from the full numerics.
  }
\end{figure}

In the following, our aim is to relate the contributions
\eqref{eq:admittancecontributions} to quantities that can be
calculated in the absence of the ac drive \cite{noteonnumerics}.  The
regimes where the different contributions are relevant depend on the
particular values of the phase difference, frequency and
temperature. The main results are summarized in
Fig.~\ref{fig:phasedepregimes}, which shows the phase dependence of
the admittance in different regimes of frequencies and temperatures.

{\em Low frequency:} For frequencies satisfying $\Gamma \ll \omega
\ll E_g(\varphi)$, the superconducting correlations follow
the time-dependent phase difference, but the electron distribution is
driven out of equilibrium.  In this limit, the time-dependent
supercurrent \eqref{eq:sccurrent} yields
Eq.~\eqref{eq:josephsoninductance}, i.e., $Y_{\rm
  sc}\simeq-2i\partial_\varphi I_S(\varphi)/\omega$.  Since
the supercurrent $I_S$ decays exponentially as the temperature increases,
this contribution to reactance becomes unimportant at high temperatures
($T\gtrsim{}10E_T$), unless the frequency is very low.

The second major contribution to reactance comes from a dynamic
variation of the population, as given by $Y_{\rm dy}$.
For this, the time dependent component
$\hat{h}_{01}\equiv{}h_L'+\tz h_T'$ of the distribution function needs to be
solved from a kinetic equation. Assuming again simple time dependence
for the superconducting correlations, the first harmonic of the
Usadel kinetic equation reads (cf.~\cite{virtanen2010-tom})
\begin{subequations}
  \label{eq:kineticeqs}
  \begin{gather}
    D\partial_x \cdot ({\cal D}_L \partial_x h_L'-{\cal T}\partial_x h_T'+j_S
    h_T')=
    \\\nonumber
    \frac{iA_0}{2}(j_S+\partial_x {\cal T})(h_{11}-h_{00})+i(\omega-2i\Gamma)N h_L'
    \,,
    \\
    D\partial_x \cdot ({\cal D}_T \partial_x h_T'+{\cal T}\partial_x h_L'+j_S
    h_L')=0.
  \end{gather}
\end{subequations}
Here ${\cal D}_{L/T}$ are the spectral heat/charge diffusion
coefficients, ${\cal T}$ is an anomalous kinetic coefficient and $N$
is the local density of states. These quantities are related to the
equilibrium Retarded Green's function, $\hat{g}^R(E) = g(E)
\hat{\tau}_3 + f(E)\hat{\tau}_\uparrow -
\tilde{f}(E)\hat{\tau}_\downarrow$, e.g.,
$j_S=\Im j_E$, $j_E=\frac{1}{2i}\Tr\hat{\tau}_3\hat{j}^R_{00}$,
$N=\Re g$, and ${\cal D}_T=\frac{1}{2}[1+|g|^2+|f|^2/2+|\tilde{f}|^2/2]$.

At low frequencies, Eqs.~\eqref{eq:kineticeqs} should be solved
with Andreev reflection boundary conditions amounting to $h_T'=0$ and
$\partial_x \cdot h_L' \hat{n}=0$ at the two NS interfaces. The resulting
function $h_T'$ is in the vector potential gauge finite but small, and as a first
approximation we can disregard it. Moreover, gradients of $h_L'$
are small due to Andreev reflection, and we get a fairly good estimate for
the average $h_L'$ by averaging Eqs.~\eqref{eq:kineticeqs} over the
normal-metal junction, defining $\avg{\cdot} = \int_0^L dx
\cdot/L$. As a result, we get
\begin{equation}
  \hat h_{01}
  \approx
  \avg{h_L'}
  \approx
  \frac{A_0}{2}
  \frac{j_S
  (h_{11}-h_{00})}{(\omega-2i\Gamma) \langle N \rangle}.
\end{equation}
Substituting this to Eq.~\eqref{eq:dycurrent} and assuming
$\omega\lesssim{}E_T$,
\begin{align}
  \label{eq:dynamic}
  Y_{\rm dy}
  &
  \equiv \frac{-iG_N}{\omega-2i\Gamma}
  \frac{E_T}{T}Q(\varphi,T)
  \\
  &\approx
  \frac{-iG_N}{\omega-2i\Gamma}\int_{-\infty}^\infty dE
  \frac{j_S^2}{4T\avg{N}\cosh^2[E/(2T)]} .\nonumber
\end{align}
This is similar to a correction to the dc
conductance described by Lempitskii \cite{lempitskii1983-sos}.
Its origin can be understood as follows~\cite{chiodi2010-mro}:
the current is carried by a dense spectrum $\{\epsilon_n(\varphi)\}$
of discrete bound states with populations $f_n$,
$j(\varphi)=\sum_n (\partial_\varphi\epsilon_n) f_n(\epsilon_n)$.
With ac bias one finds $\delta j/\delta V \propto
\sum_n \partial_\varphi^2\epsilon_n f_n/i\omega + \sum_n (\partial_\varphi\epsilon_n)^2\partial_\epsilon f_n/i\omega$,
where the first term is equivalent to $Y_{\rm sc}$ and the second
one to $Y_{\rm dy}$, as
$j_S\sim{}\sum_n\delta(E-\epsilon_n)\partial_\varphi\epsilon_n\sim{}N\partial_\varphi\epsilon$.

The dynamic contribution is purely dissipative and constant for
$\omega \ll \Gamma$, contains both reactive and dissipative
components for $\omega \approx \Gamma$, and becomes purely reactive
and decays for $\omega \gg \Gamma$, as visible in
Fig.~\ref{fig:phasedepregimes}. In general, $\Gamma$ also depends on
the phase difference \cite{heikkila2009-pse}, so that the
phase-dependent response at frequencies of the order of the inelastic
scattering rate may be quite complicated. On the other hand,
Eq.~\eqref{eq:dynamic} offers a way to probe such phase-dependent
scattering rates via an admittance measurement.

\begin{figure}
   \includegraphics{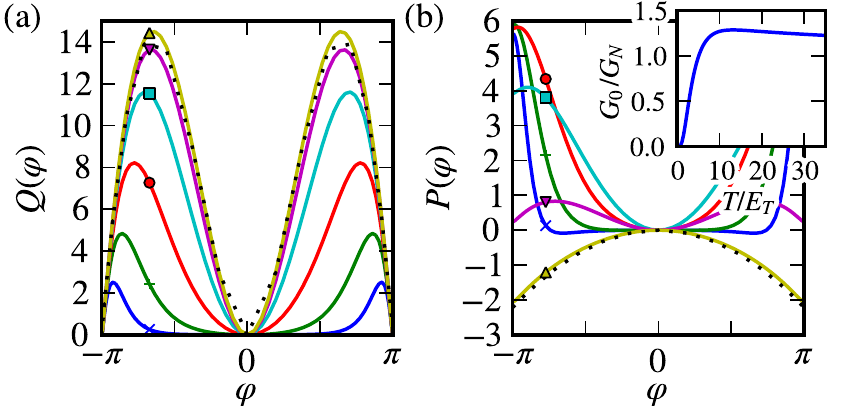}
   \caption{
     \label{fig:QP}
     (Color online): (a) Function $Q(\varphi,T)$ (see Eq.~\eqref{eq:dynamic}) describing the
     phase-dependent dynamic contribution to the reactive response and
     (b) function $P(\varphi,T)$ (Eq.~\eqref{eq:dissipative}) describing the phase dependence of the
     dissipative part of the admittance at low frequencies. From
     top to bottom in (a) and marked with symbols in (b):
     $T/E_T=16(\vartriangle), 8(\triangledown), 4(\square), 2(\circ), 1(+), 0.5(\times)$. The dotted lines represent the analytic
     high-temperature approximations to which $Q$ and $P$ tend for
     $T \gg E_T$.  The inset of (b) shows the temperature
     dependence of $G_0(T)$.
   }
\end{figure}

The function $Q(\varphi,T)$ is shown in Fig.~\ref{fig:QP}(a) for
different temperatures. At $T \gg E_T$ the response can be fitted with
the function $Q(\varphi)\approx
8.9\bigl(|\saw(\varphi)|-0.32\sin^2(\varphi/2)\bigr)|\sin(\varphi)|$ where
$\saw(\varphi)=(\varphi+\pi\mod2\pi)-\pi$.  At low temperatures, $Q(\varphi,T)$
is suppressed for phases at which $T < E_g(\varphi)$. Note that $Q$
is positive definite, has almost a double periodicity compared to the
kinetic inductance term, and has a minimum around $\varphi=0$, where the
kinetic inductance is at maximum.

The dissipative part of the impedance originates from two additional
sources: the
quasiparticle part $Y_{\rm qp}$, and, importantly at low temperatures,
a part of the AC supercurrent oscillating in phase with the voltage.
The quasiparticle contribution is easiest to derive in a gauge where
the vector potential vanishes. Then, $h_T'$ in Eqs.~\eqref{eq:kineticeqs}
has the boundary conditions
$h_T'(x=\pm{}L/2)=\pm\frac{i\delta\phi}{4}(h_{11}-h_{00})$,
and solving Eq.~\eqref{eq:kineticeqs} assuming $h_L'\approx0$ yields
\begin{align}
  Y_{\rm qp} + \Re Y_{\rm sc}
  &\simeq G_N
  \int_{-\infty}^\infty  \dd{E} \frac{K_0(E)}{
    4 T \cosh^2[E/(2T)]}
  \,,
  \\
  &
  \equiv G_0(T) + \frac{G_N E_T}{T}P(\varphi,T),
  \label{eq:dissipative}
\end{align}
where we assumed $\omega\lesssim{}E_T$, and defined $G_0(T)$ as
the $\varphi=0$ value [$P(0,T)=0$].  The kernel is
$K_0\approx\avg{{\cal D}_T^{-1}}^{-1} - \Re \partial_\varphi
j_E\sim{}\avg{N^2+\frac{1}{4}|f+\tilde{f}^*|^2}$ \footnote{ The
  latter approximation is the one used in \cite{virtanen2010-tom} and
  captures the minigap correctly, whereas the former gives a better
  approximation to the integral, but is inaccurate at energies
  $E\lesssim{}E_g(\varphi)$.  }.  It describes the spectrum of
excitations in the junction available for receiving energy --- this has
a minigap $E_g(\varphi)$, so that for $\omega, T < E_g(\varphi)$ all
dissipation vanishes. Note that the appearance of the minigap is
related to the presence of the AC supercurrent contribution:
$Y_{\rm qp}\ge{}G_N$ and is similar to the usual proximity-enhanced
conductance.

The functions $P(\varphi,T)$ and $G_0(T)$ are shown in
Fig.~\ref{fig:QP}(b).  At low temperatures, the temperature and phase
dependence shows a clear signature of the presence of a minigap in the
density of states; note that this also applies to the $Y_{\rm dy}$
contribution.  The dissipation is concentrated at phase differences
close to $\pi$ where the minigap is small. In the high-temperature
limit, the dissipative term consists of a phase independent
contribution $G_0(T)$ and the phase-dependent part has a simple
phase and temperature dependence, $-0.23\saw(\varphi)^2G_N E_T/T$.

{\em High frequency:} When the frequency becomes of the order of the
Thouless energy $E_T$, the above semi-adiabatic expressions break down
as the spectral quantities become frequency dependent.  We can
however construct approximations also in this limit (for full expressions,
see Appendix).  The supercurrent contribution is
fairly approximated by
\begin{equation}
  \begin{split}
    Y_{\rm sc} &
    \approx
    \frac{\sigma_N S}{4\omega}
    \int_{-\infty}^\infty\dd{E}
    \bigg\{\partial_\varphi[j_E(E)+j_E(E+\omega)]^*
    h_{00}(E)
    \\
    &\qquad
    -\partial_\varphi[j_E(E)+j_E(E+\omega)] h_{11}(E)\bigg\},
  \end{split}
  \label{eq:sc-highfreq}
\end{equation}
and the quasiparticle part of the impedance by
\begin{align}
  \label{eq:iqp-highfreq}
  Y_{\rm qp}
  &\simeq
  G_N\int_{-\infty}^\infty dE  K(E,\omega)
  \frac{h_{11}(E)-h_{00}(E)}{2\omega}
  \,,
  \\
  &K(E,\omega)
  =
  \frac{1}{2}
  \Bigl\langle
  \bigl[
  1 + g(E) g(E+\omega)^*
  \\\notag
  &\;
  + \frac{1}{2}f(E)f(E+\omega)^*
  + \frac{1}{2}\tilde{f}(E)\tilde{f}(E+\omega)^*
  \bigr]^{-1}
  \Bigr\rangle^{-1}
  \,.
\end{align}
The dynamic contribution $Y_{\rm dy}$ can be neglected for
$\omega\gg{}E_T$.  The above is compared to full numerical solutions in the
top panels in Fig.~\ref{fig:phasedepregimes}.

At high temperatures, $Y_{\rm sc}$ is exponentially suppressed even for
high frequencies, similarly as the equilibrium
supercurrent. Consequently, $Y_{\rm qp}$ dominates for
$\omega,T\gg{}E_T$ (see Fig.~\ref{fig:phasedepregimes}). At low
temperatures, both the supercurrent and quasiparticle contributions
are important.

\begin{figure}
  \includegraphics{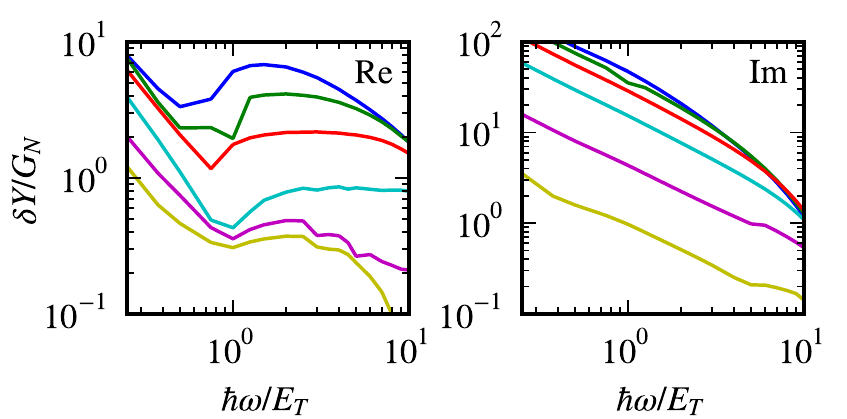}
  \caption{\label{fig:highwphi}
    (Color online): Amplitude
    $\delta{}Y=\max_\varphi [\Re/\Im] Y(\varphi) - \min_\varphi [\Re/\Im] Y(\varphi)$ of the
    phase-dependence in the nonadiabatic dissipative (Re)
    and reactive (Im) admittance of a long SNS junction, obtained by
    solving the time-dependent Usadel equations numerically.  The
    lines correspond to temperatures $T/E_T=0.5, 1, 2, 4, 8, 16$
    (top to bottom) and were calculated with $\Gamma=0.05 E_T$.
  }
\end{figure}

The amplitude of the phase dependence in the admittance is illustrated
in Fig.~\ref{fig:highwphi}, up to high frequencies. As the frequency
increases, the proximity-induced phase dependence in the reactive and
the dissipative components decays, as the admittance approaches the
constant normal-state value, $Y(\omega,\varphi)\to{}G_N$. Similarly,
the phase dependence vanishes as the temperature increases. At low
frequencies, on the other hand, the reactive component diverges due to
the Josephson inductance, and the dissipative component is dominated
by $Y_{\rm dy}$.

In a recent experiment probing directly the ac admittance of SNS
junctions \cite{chiodi2010-mro}, it was found that at frequencies
$\Gamma\ll\omega\ll{}E_T$ (semi-adiabatic limit), the reactive
response follows closely our prediction consisting of the sum of the
Josephson inductance and the dynamical correction
\eqref{eq:dynamic}. However, both the dissipative contribution and the
dependence at high frequencies $\omega \gtrsim E_T$ are different: in
\cite{chiodi2010-mro}, the dissipative contribution is directly
related to the reactive contribution, and moreover the amplitude of
phase oscillations in susceptibility $\delta\chi=i\omega \delta Y$ decays as
frequency increases. The characteristic frequency scale for this was
found to be temperature independent and of the order of the Thouless energy. In
contrast, in this frequency range our Usadel model predicts
$\delta\chi\sim\text{const.}$ --- however, it is for example possible that
the simple relaxation time approximation does not include all
interaction mechanisms playing a role in the experiment.


Finally, we remark that our work amounts essentially to deriving the
parameters for the Resistively Shunted Junction (RSJ) model of SNS
junctions \cite{tinkham1996-its}: there, the Josephson inductance resulting
from the supercurrent term should be modified to include the
nonadiabatic correction, Eq.~\eqref{eq:dynamic}, and the shunt
resistor describing the dissipation in the junction should be replaced
by the dissipative terms presented in Eqs.~(\ref{eq:dynamic},\ref{eq:dissipative}).


In conclusion, we have described the frequency-dependent admittance of
diffusive superconductor-normal metal-superconductor junctions and
shown how the simple adiabatic Josephson inductance picture is
modified once the frequency is increased. Besides studying the
dynamics of the system, the detailed frequency dependence can be used
to study directly the inelastic scattering rates. Our results are also
relevant for devices utilizing high-frequency properties of SNS
junctions, such as those used in metrology and radiation detection.

We thank H.~Bouchiat, S.~Gueron, K.~Tikhonov, and M.~Feigelman for
discussions that in part motivated this work, and CSC (Espoo) for
computer resources. This work was supported by the Academy of Finland,
the ERC (Grant No. 240362-Heattronics), the Spanish MICINN (Contract
No. FIS2008-04209), and the Emmy-Noether program of the DFG.

\bibliography{snsimpedance}

\appendix

\section{On approximations}

Although one cannot solve the time-dependent linear-response Usadel
equations analytically in closed form, the admittance can be obtained
with quantitative accuracy if the solution at equilibrium is known.
The approximation procedure is outlined in the main text.  Below, we
show intermediate results before taking the $\omega\to0$ limit, and
demonstrate that the results agree with the full numerical approach
which solves the ac Usadel equation exactly (see the Supplementary
information of Ref.~\onlinecite{virtanen2010-tom} for details of the
numerics).

Following the procedure outlined in the main text, the following
approximations can be derived:
\begin{widetext}
\begin{subequations}
\label{eq:Yapprox}
\begin{align}
  Y &
  = \sigma S \int_{-\infty}^\infty dE\: \tilde{Y}(E)
  = \sigma S \int_{-\infty}^\infty dE\: [\tilde{Y}_{\rm sc}(E)+\tilde{Y}_{\rm dy}(E)+\tilde{Y}_{\rm qp}(E)]
  \,,
  \\
  \tilde{Y}_{\rm sc}(E) &\simeq
  \frac{-1}{2\omega}\Bigl(\frac{\partial_\phi[j_{E}(E)+j_{E}(E+\omega)]}{2}h(E+\omega)-\frac{\partial_\phi[j_{E}(E)+j_{E}(E+\omega)]^*}{2} h(E)\Bigr)
  \,,
  \\
  \tilde{Y}_{\rm dy}(E) &\simeq
  \frac{1}{8\omega}\frac{i}{\omega- 2i\Gamma}\frac{(j_{E}(E)-j_{E}(E+\omega)^*)^2}{\frac{1}{2}\langle{}g(E)+g(E+\omega)^*\rangle}(h(E+\omega)-h(E))
  \,,
  \\
  \tilde{Y}_{\rm qp}(E) &\simeq
  \frac{1}{4\omega}\bigl\langle[1 + g(E) g(E+\omega)^* + \frac{1}{2} f(E)f(E+\omega)^* + \frac{1}{2}\tilde{f}(E)\tilde{f}(E+\omega)^*]^{-1}\bigr\rangle^{-1}(h(E+\omega)-h(E))
  \,.
\end{align}
\end{subequations}
\end{widetext}
Above, $h(E)=\tanh[E/(2T)]$, and the energy-dependent quantities
$j_E(E) = \frac{-i}{2}\Tr[\hat{\tau}_3\hat{g}^R\nabla\hat{g}^R]$ and
$\hat{g}^R(E) = g(E) \hat{\tau}_3 +
f(E)\frac{\hat{\tau}_1+i\hat{\tau}_2}{2} -
\tilde{f}(E)\frac{\hat{\tau}_1-i\hat{\tau}_2}{2}$ are obtained from
the equilibrium Usadel equations \cite{belzig1999-qgf,usadel1970-gde}, which are much faster to solve than the full ac equations.

\begin{figure*}
  \includegraphics{fig1}\includegraphics{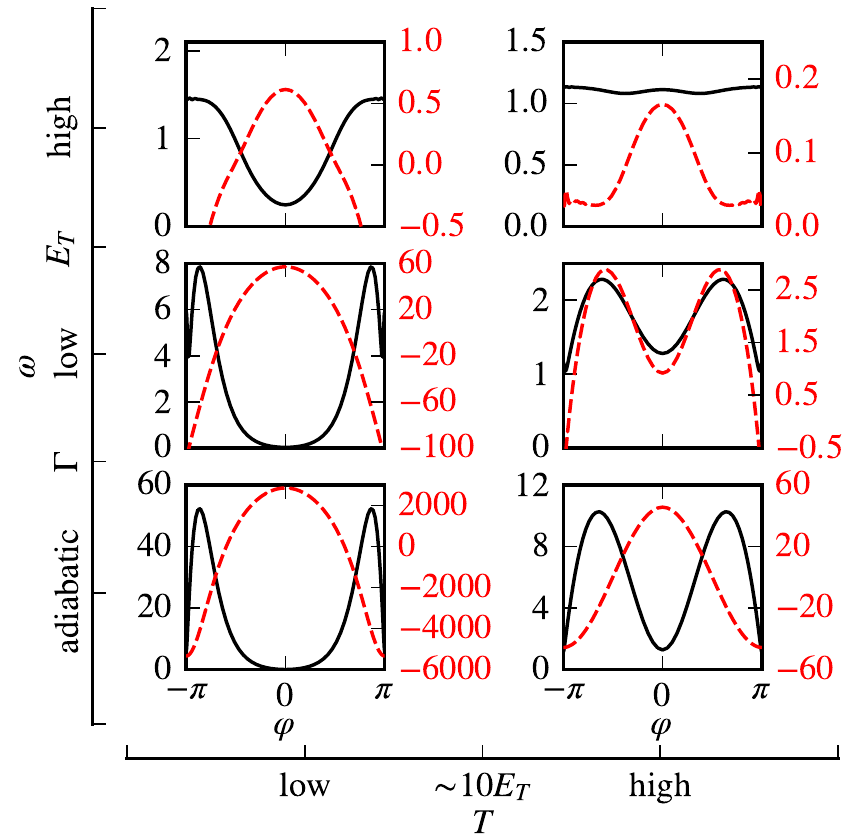}
  \caption{
    \label{fig:phasedepregimes-an}
    Left: Fig.~1 in the main text.
    Right: Fig.~1 with all data computed from Eqs.~\eqref{eq:Yapprox}.
  }
\end{figure*}

\begin{figure*}
  \includegraphics{fig3_2d}\includegraphics{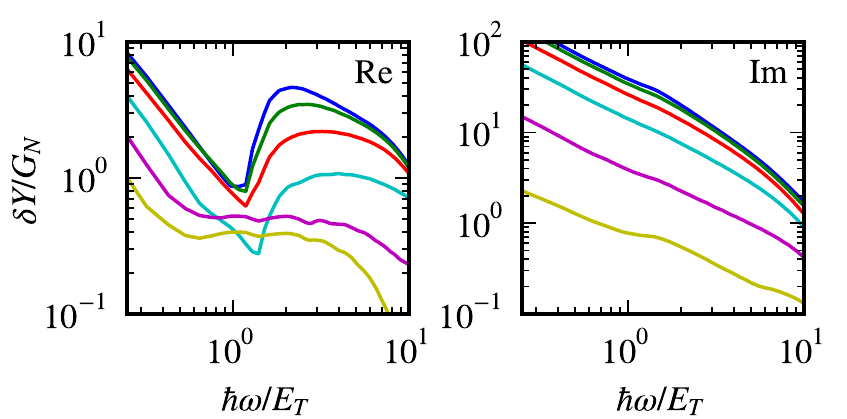}
  \caption{\label{fig:highwphi-an}
    Left: Fig.~3 in the main text.
    Right: Fig.~3 with all data computed from Eqs.~\eqref{eq:Yapprox}.
  }
\end{figure*}

\begin{figure*}
  \includegraphics{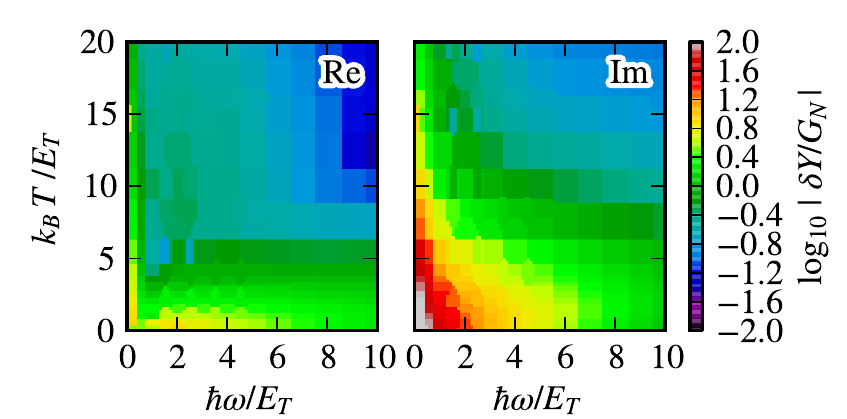}\includegraphics{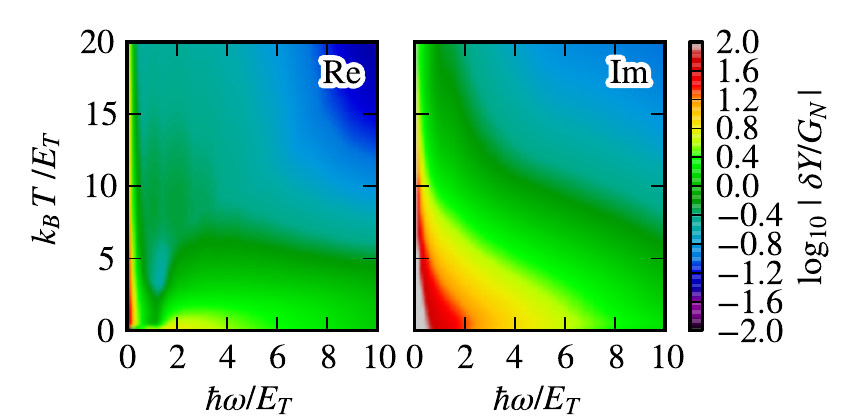}
  \caption{\label{fig:highwphi-3d-an}
    As Fig.~3 in the main text, but expressed as a color plot.
    Left: numerically computed admittance oscillations.
    Right: admittance oscillations from Eqs.~\eqref{eq:Yapprox}.
  }
\end{figure*}

Figures
\ref{fig:phasedepregimes-an}, \ref{fig:highwphi-an}, and
\ref{fig:highwphi-3d-an} illustrate that the above approximations
reproduce all qualitative features visible in the fully numerical
solution, and are quantitatively accurate in a large part of the
parameter regime we are interested in. In particular, the deviations between the two approaches are miniscule for frequencies lower than $E_T$. At large frequencies there are clear quantitative differences, but the qualitative phase and frequency dependence is the same, and the resulting admittances are of the same order of magnitude. This demonstrates that to a fair accuracy the ac admittance can be well described by using Eqs.~\eqref{eq:Yapprox} and standard solvers for the equilibrium Usadel equation. Besides that, Eq.~\eqref{eq:Yapprox} provides a possibility for making analytical estimates for the admittance contributions.

\end{document}